\documentclass[fleqn,showpacs,twocolumn,amsmath,amssymb]{revtex4-1}

\usepackage{forloop,ifthen} 
\usepackage{multirow} 
\usepackage{ esint }
\usepackage{mathtools}
\usepackage{amsmath} 
    \usepackage{graphicx}
    \usepackage{transparent}
    \usepackage{fancybox}  
    \usepackage{color}
    \usepackage{dcolumn}
    \usepackage{bm}
    \usepackage{xspace}
    \usepackage{float}
    \usepackage[colorlinks=true,     linkcolor={red!70!black},
    citecolor={green!50!black}, urlcolor={blue!50!black}, pdfborder={0 0 0}]{hyperref}
	\usepackage{calc}
   	
\usepackage[usenames,dvipsnames]{xcolor}
\usepackage{ amssymb }

\newcommand{\suppress}[1]{}

\newcommand{\bea}[1]{\begin{eqnarray}\label{#1}}
\newcommand{\eea}{\end{eqnarray}}

\makeatletter
\newcommand{\eqlab}[1]{\refstepcounter{equation}\label{#1}\textup{\tagform@{\theequation}}}
\makeatother

\begin{document}

\title{Structures far below sub-Planck scale in quantum phase-space through superoscillations}

\author{Maxime Oliva, Ole Steuernagel}
\affiliation{School of Physics, Astronomy and Mathematics, University of Hertfordshire, Hatfield,
  AL10 9AB, UK}
\date{\today}

\begin{abstract}

  In 2001, Zurek derived the generic minimum scale $a_{Z}$ for the area of structures of
  Wigner's quantum phase distribution. Here we show by construction, using
  superoscillatory functions, that the Wigner distribution can locally show regular spotty
  structures on scales much below Zurek's scale $a_{Z}$. The price to pay for the presence
  of such structures is their exponential smallness. For the case we construct there is no
  increased interferometric sensitivity from the presence of patches with superoscillatory
  structure in phase-space.
  
\end{abstract}

\maketitle
\section{Introduction}

Based on the concept of interferences in phase-space~\cite{Schleich_01}, Zurek established
that the minimum scale $a_{Z}$ for the area of structures of quantum phase distributions
can, for one-dimensional quantum systems, be as small as $a_{Z} \approx \hbar^{2}/A$,
where $A$ is the action representing the area of support of a system's Wigner
distribution~\cite{Wigner_PR32}. This was surprising~\cite{Zurek_01}, since Heisenberg's
uncertainty principle was interpreted to limit the area of spots in phase-space to
approximately $\hbar/2$.

When a bandwidth-limited signal contains segments which oscillate faster than what its
spectrum suggests, it ``superoscillates"~\cite{Berry_WS94}.

Superoscillations have first been noticed in physics in the framework of weak measurement
by Aharonov et al.~\cite{Aharonov_PRL90}, and then studied in detail by Aharonov et
al.~\cite{Aharonov_JPA11}, Berry et al.~\cite{Berry_WS94,Berry_JPA06}, and
others~\cite{Kempf_JMP05}. They have since been used experimentally, for example in
super-resolution micro\-scopy~\cite{Dennis_NATMat12}.

Translated to quantum phase-space: Wigner distributions can show regular spotty structures
much below Zurek's scale $a_{Z}$; but such states cannot obviously be exploited for higher
resol\-ution in measurements.

\section{Zurek's fundamental phase-space tiles}
The Wigner distribution of a ``Schr\"odinger's cat" state of squeezed states $G(x,p)=(\pi
\hbar)^{-1} e^{-x^{2}/\xi^{2} - \frac{p^{2}\xi^{2}}{\hbar^{2}} }$, with squeezing
parameter~$\xi$, is
\begin{subequations}
\begin{align}
W(x,p)=\frac{G(x-\Delta x,p)+G(x+\Delta x,p)}{2} \label{eq:compassstate}\\
 +G(x,p)\cos\left(\frac{2 p }{\hbar} \Delta x\right) \label{eq:interferenceterm}.
 \end{align}\label{eq:WSchrodingersCat}
\end{subequations}

\begin{figure}[t]
	\includegraphics[width=0.50\columnwidth]{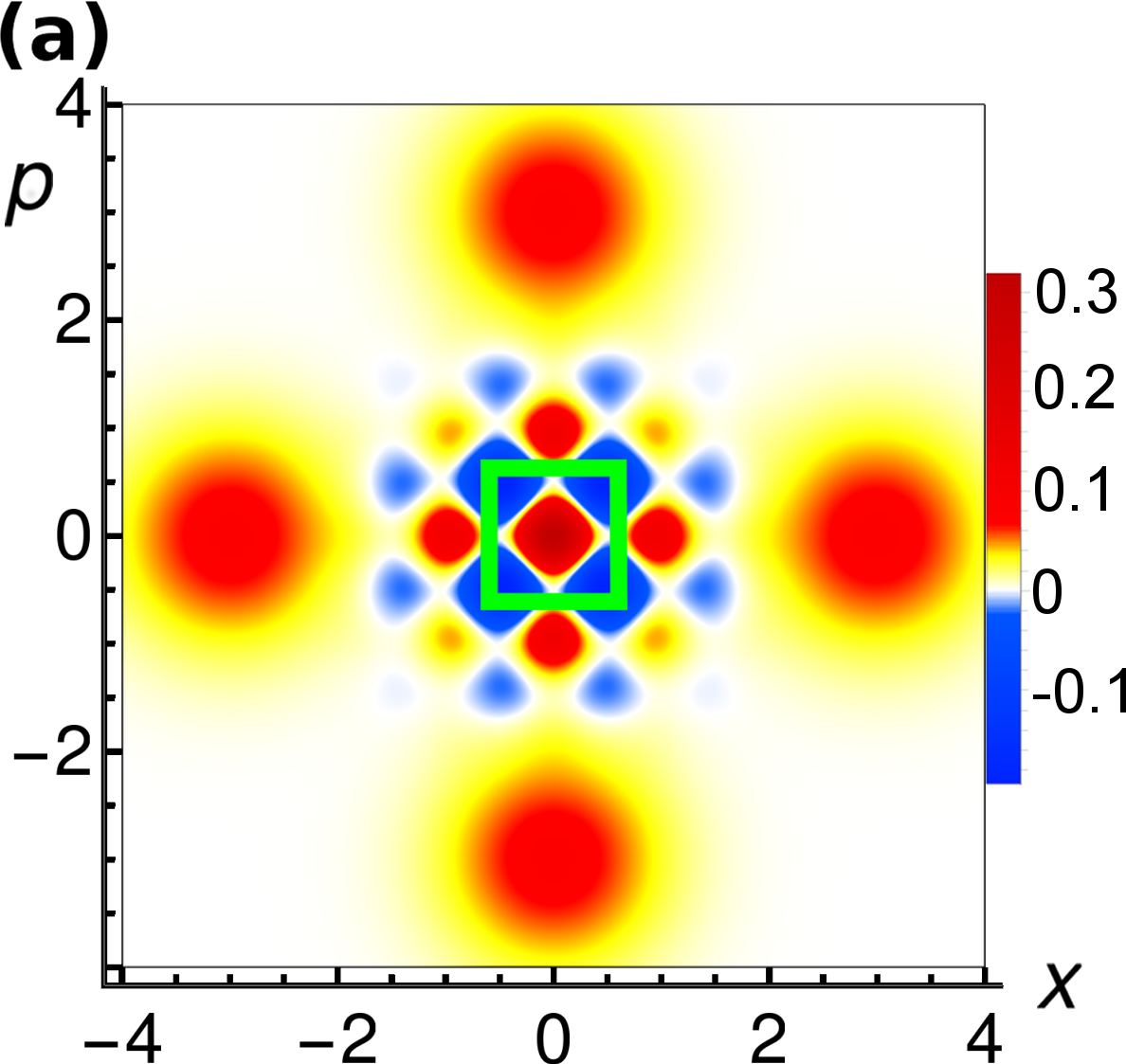}
	\includegraphics[width=0.48\columnwidth]{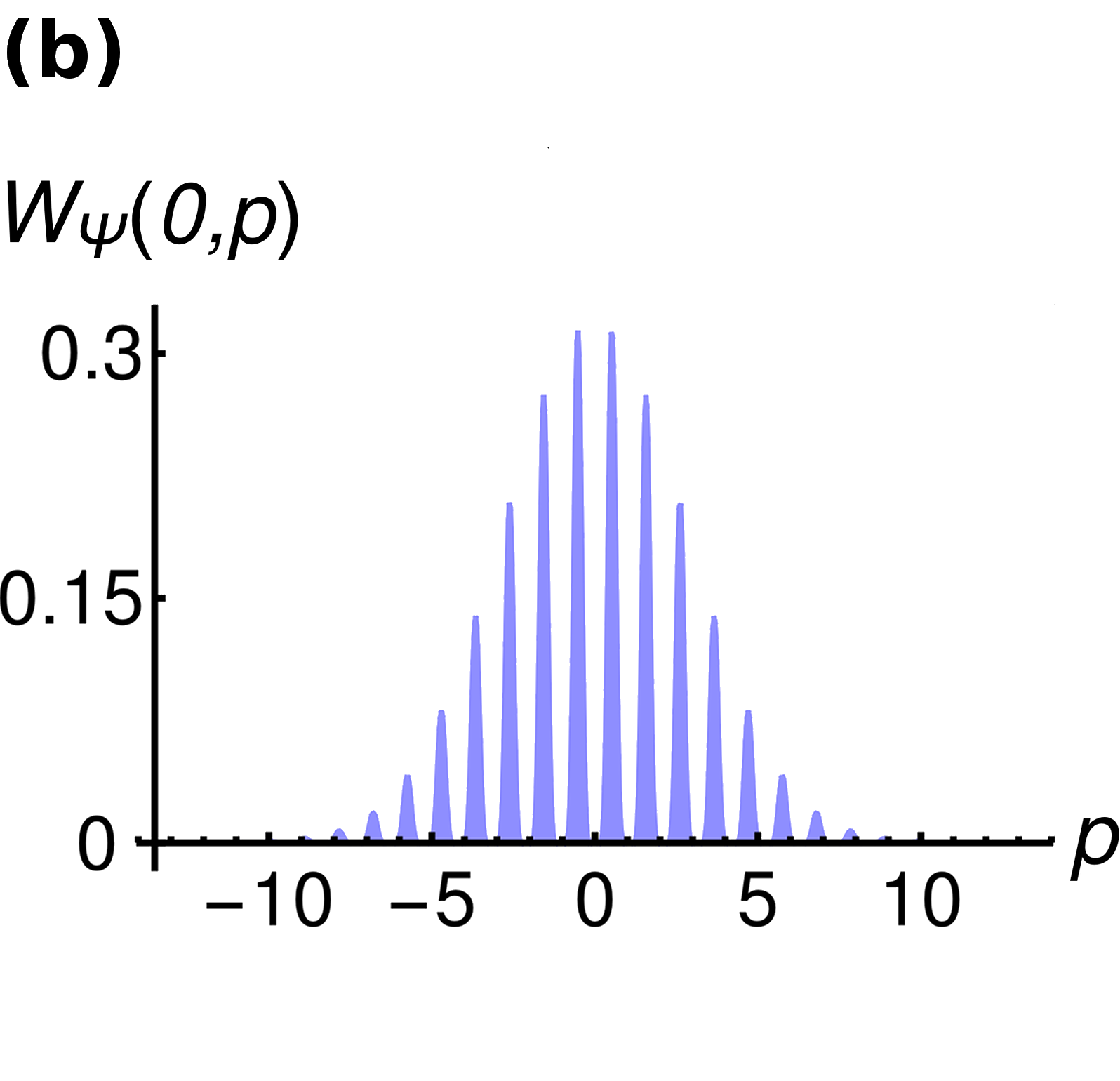}\\
	\includegraphics[width=0.48\columnwidth]{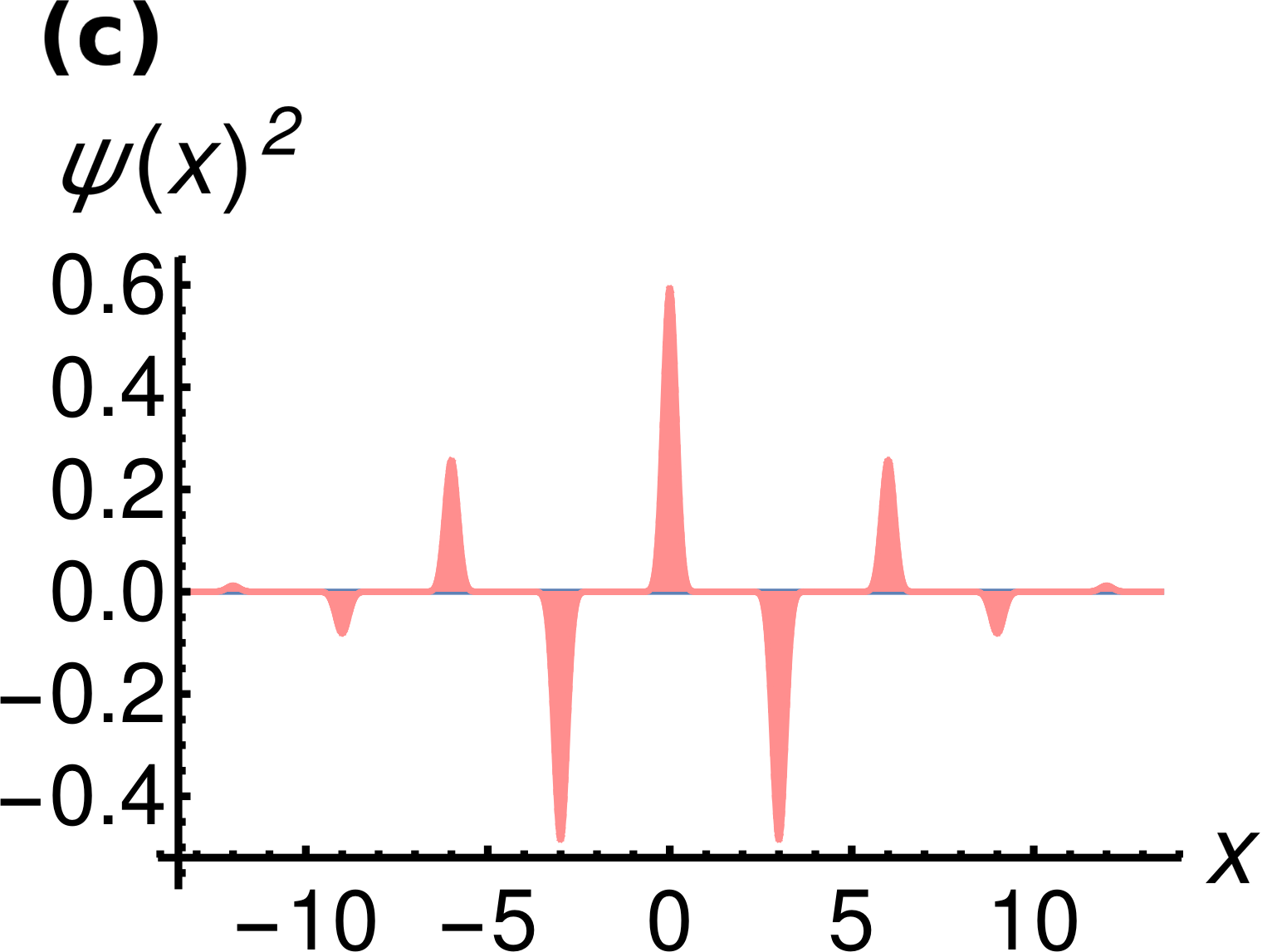}
	\includegraphics[width=0.50\columnwidth]{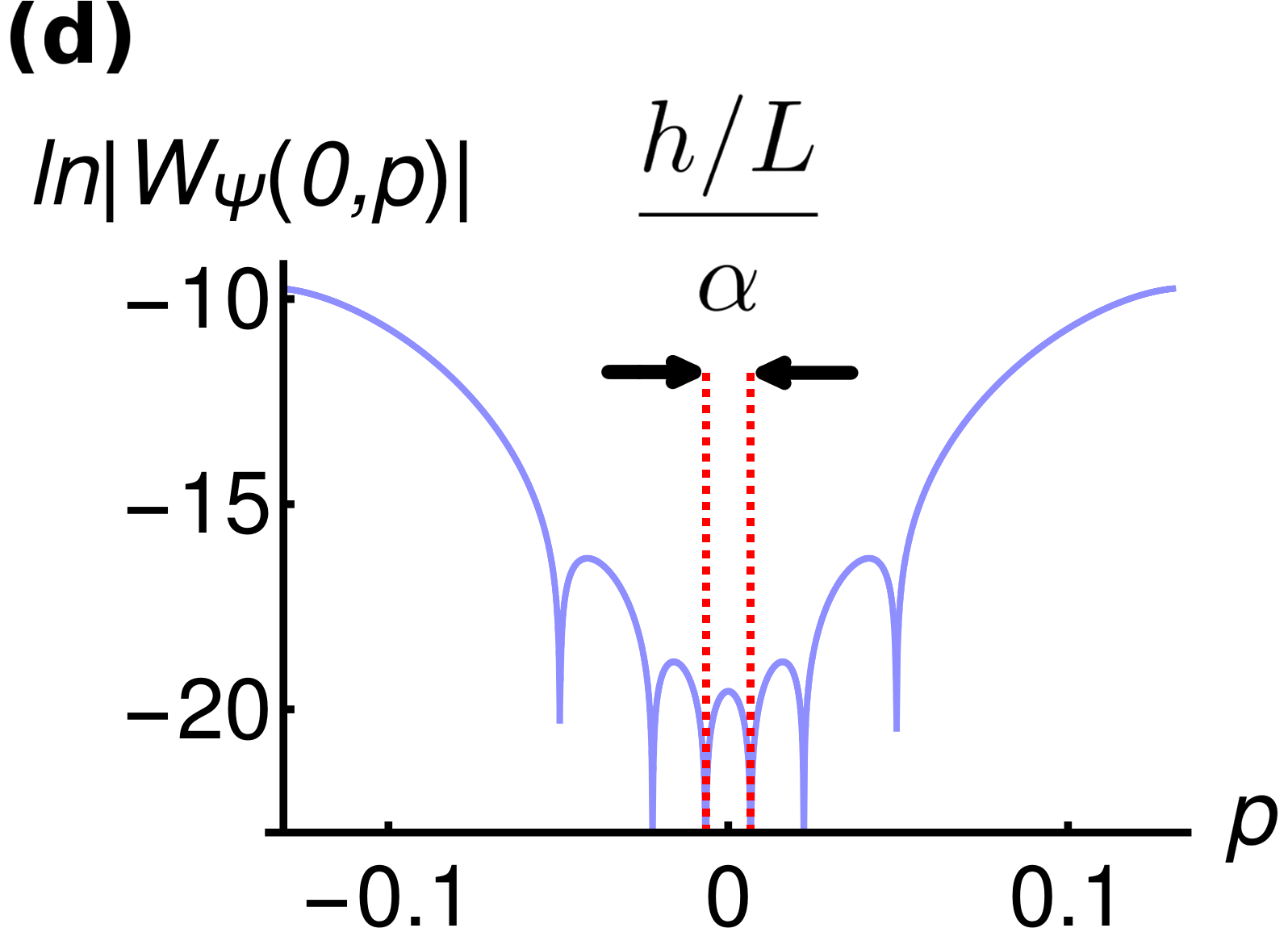}
        \caption{(Color online) {\bf(a)} shows a compass state~\cite{Zurek_01} for $L=P=6$
          and $\xi=1$ (atomic units [a.u.], $\hbar=1$, are used in all figures), the green
          frame borders its Zurek tile with area~$a_{Z}$. {\bf(b)}, {\bf(c)} and {\bf(d)}
          are obtained using $N=8$, $\alpha =10$, $\xi=\frac{1}{4}$, $\Delta x=3$ and
          $\hbar=1$. {\bf(c)} shows the squared wave function $\Psi(x)^{2}$ and its sign
          changes due to the complex coefficients in $\Psi(x)$
          [Eq.~(\ref{eq:SO_wavefunction})]. {\bf(b)} shows $W_{\Psi}(0,p)$, and {\bf(d)}
          the logarithm of $|W_{\Psi}(0,p)|$ near the origin, in a panel of width
          $h/L\approx 0.26$, in which the two red dashed lines bracket superoscillatory
          structure of length $\frac{h/L}{\alpha}=\frac{h/L}{10}$ [see
          Eq.~(\ref{eq:sub-zurekscale})].}
\label{fig1}
\end{figure}

Zurek's compass state is a coherent sum of two Sch\"odinger's cat states 
rotated by $\pi/2$ with respect to each other. Starting from such compass 
states, Fig.~\ref{fig1}~{\bf(a)}, Zurek showed that \emph{``Wigner functions can, 
and
generally will, develop phase-space structures on scales as small as,
but not generally smaller than"}~\cite{Zurek_01}

\begin{eqnarray}
\label{eq:zurekscale}
 a_{Z} = \frac{h}{P} \times \frac{h}{L} = \frac{h^{2}}{A}.
\end{eqnarray}

Here, $P$ and $L$ are the phase-space distances between the squeezed states along the
momentum and position axes respectively.

The Zurek scale~$a_Z$ is, e.g., the phase-space area of one fundamental
tile~\cite{Zurek_01} (``Zurek tile") associated with a compass state, highlighted in
Fig.~\ref{fig1}~{\bf(a)}.

\section{The superoscillating cross-state\label{sec_SuperOscilllatingCrossState}}

\begin{figure}[t]
 \includegraphics[width=0.45\textwidth]{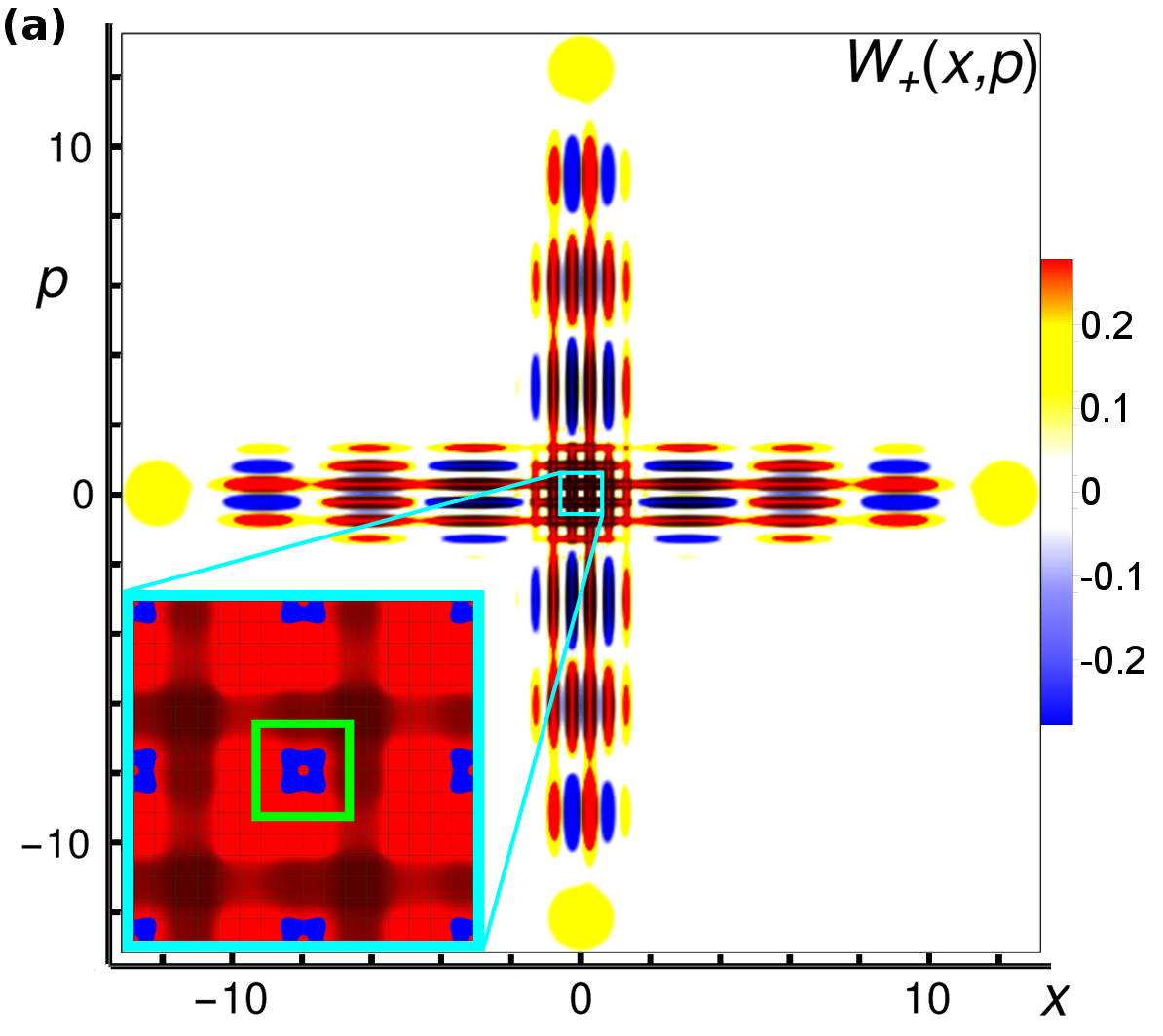}
\\
 \includegraphics[width=0.225\textwidth]{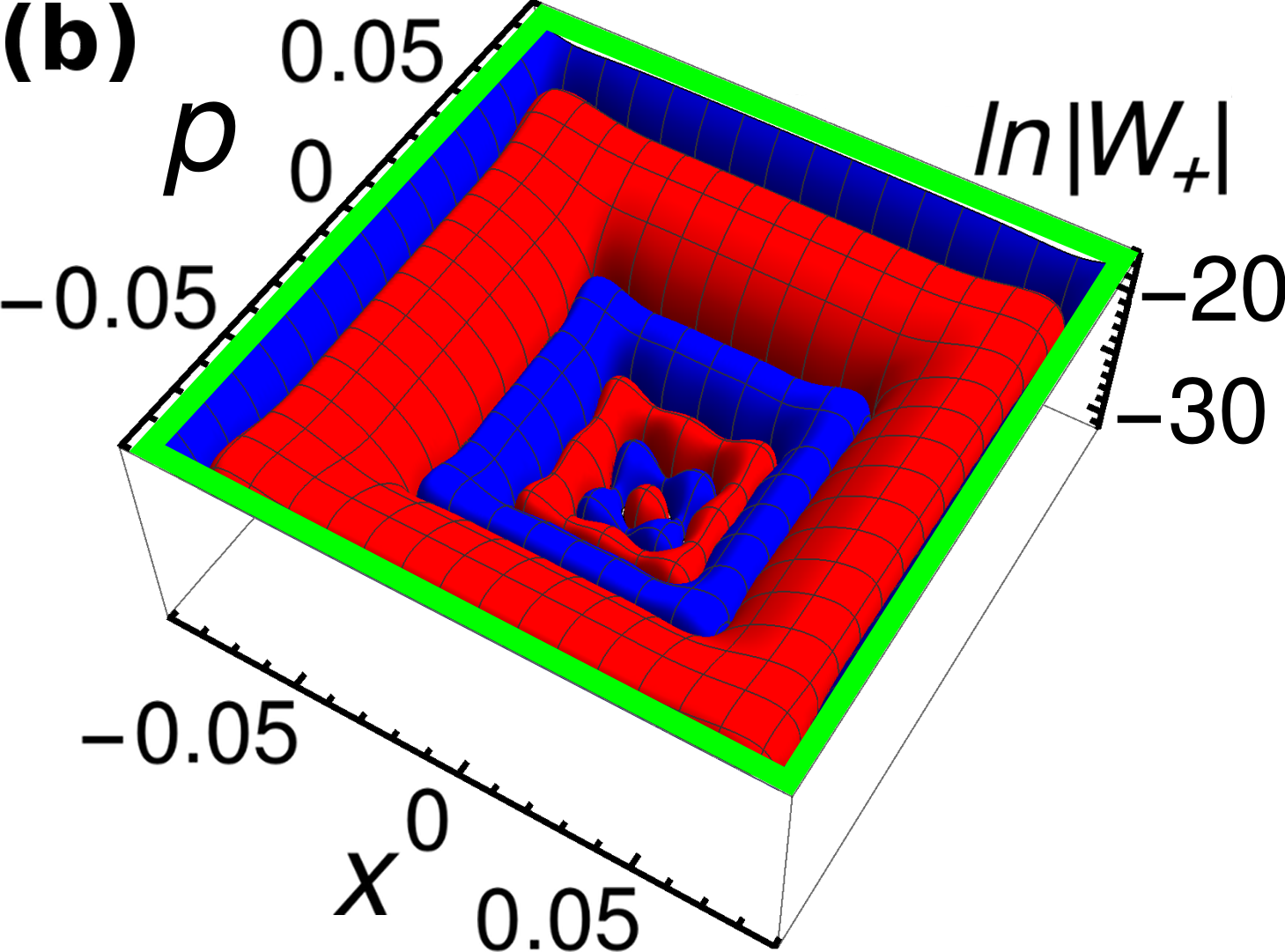}
     \includegraphics[width=0.225\textwidth]{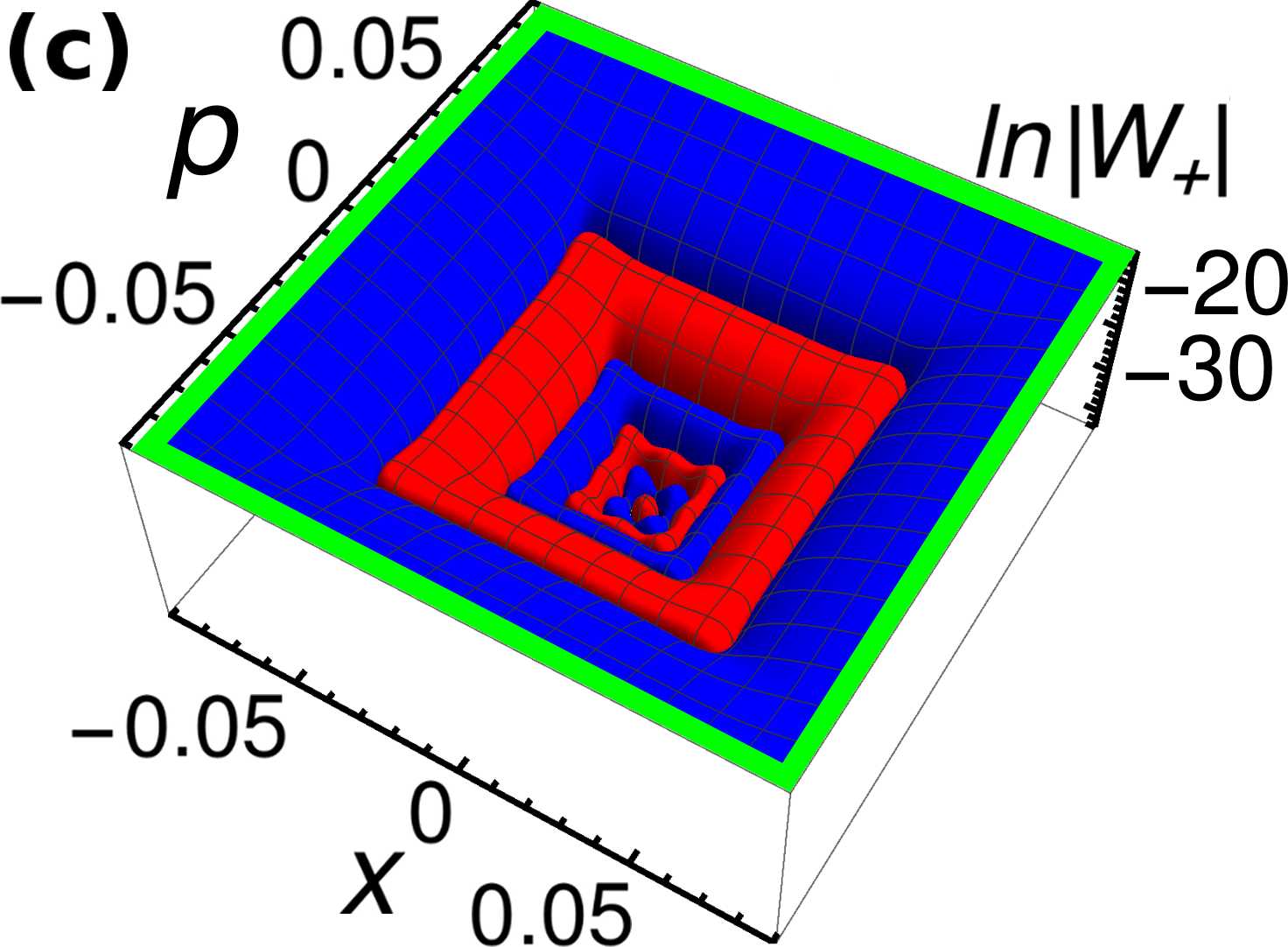}
     \caption{(Color online) \textbf{(a)} Cross-state
       $W_{+}(x,p)=[W_{\Psi}(x,p)+W_{\Psi}(-p,x)]/2$ for $N=4$, $\alpha=6$, $\xi=1$, and
       $\Delta x=6$ ($\hbar=1$, a.u.). In \textbf{(a)}'s inset and panels
       \textbf{(b)} and \textbf{(c)}, red (positive) and blue (negative) regions depict
       superoscillatory structures contained within Zurek tiles (green
       frames). \textbf{(b)} and \textbf{(c)} show $\ln|W_{+}(x,p)|$ and demonstrate the
       scaling with $a_{SO}$, see Eq.~(\ref{eq:sub-zurekscale}). State parameters $N=12$,
       $\xi=\frac{1}{4}$, $\Delta x=3$, with $\alpha=10$ in \textbf{(b)}, and $\alpha=16$
       in \textbf{(c)}, respectively.}

\label{fig2}
  \end{figure}

Inspired by Zurek's compass state, we construct a ``cross-state" featuring
superoscillations in quantum phase-space. This state features small patches with regu\-lar
structures on scales much smaller than $a_{Z}$, Fig.~\ref{fig2}.

We use the superoscillating function~\cite{Aharonov_JPA11,Berry_WS94,Berry_JPA06}
\begin{eqnarray}
\label{eq:definitionSO}
f(x)=\left(\cos(x)+i \alpha \,\sin(x)\right)^{N}, \; \alpha>1, \;  N \in  \mathbb{N}.
\end{eqnarray}

For $\alpha =1$, $f(x)=e^{i N x}$ is a regular plane wave. For $\alpha >1$ and $N\gg 1$,
$f(x)$ becomes superoscillatory [see Fig.~\ref{fig1}~{\bf(d)}]

\begin{eqnarray}
\label{eq:f_fourierform}
f(x)=\sum_{j=0}^{N}C_{j}(N,\alpha )e^{i(N-2j)x},
\end{eqnarray}
where $C_{j}(N,\alpha )=(-1)^{j}\binom{N}{j}(\alpha +1)^{N-j}(\alpha -1)^{j}/2^{N}$ are
the Fourier coefficients~\cite{Aharonov_JPA11}.

To map $f(x)$ of Eq.~(\ref{eq:f_fourierform}) into phase-space, we use a superposition of
suitably pairwise-displaced squeezed states $S(x)=(\pi\xi^2)^{-1/4}\,e^{-x^{2}/(2\xi^2)}$
[see Eq.~(\ref{eq:WSchrodingersCat})], to form

\begin{flalign}
\label{eq:SO_wavefunction}
\Psi(x)=\Phi_{0}(x)+\frac{1}{\sqrt{2}} 
 \sum_{\substack{j=-N/2\\
                  j \neq 0\\
                  }}^{N/2}
                  (-i)^{j} \Phi_{j}(x), &
\end{flalign}
where $\Phi_{j}(x)=K_{j} \, S(x-j\Delta x)$, $K_{j}=\sqrt{\left|D_{j}\right|/\sum_{l=0}^{N/2}D_{l}}$ and 
\[ 
  D_{j} = 
  \begin{dcases*} 
  C_{N/2} & if  $j=0,$ \\ 
 C_{N/2+j}+C_{N/2-j} & if $j\neq 0.$ $\quad \quad \quad \quad \quad \eqlab{}$
  \end{dcases*} 
\]
Here, $N$ is even and $\Psi$ contains $N+1$ spikes, see Fig.~\ref{fig1}~{\bf(c)}.  The
associated Wigner distribution~$W_{\Psi}$ contains a suitable combination of plane wave
terms [Eq.~(\ref{eq:interferenceterm})] to emulate $f(x)$ of Eq.~(\ref{eq:f_fourierform}),
see Fig.~\ref{fig1}~{\bf(d)}.

An incoherent sum of two such Wigner distributions, rotated by $\pi/2$ with respect to
each other, forms the desired cross-state $W_{+}(x,p)=[W_{\Psi}(x,p)+W_{\Psi}(-p,x)]/2$.
This balanced mixed state features superoscillatory structures within Zurek tiles, on
sub-Zurek scales [Fig.~\ref{fig2}~{\bf(a)}-inset, {\bf(b)} and {\bf(c)}].

One could use a coherent sum to form a cross, but this would lead to greater complexity in
Fig.~\ref{fig2}, which is un\-necessary to illustrate our construction.

\section{Substructures within Zurek tiles\label{sec_GenerationOfTheSuperoscillations}}
The area of sub-Planck structures of non-superoscillating states is limited by
$a_{Z}$~\cite{Zurek_01}. Now we show that regular structures on scales much smaller
than~$a_Z$ can exist.

The local expansion of Eq.~(\ref{eq:definitionSO}) around the origin has the local
superoscillatory plane wave form
\begin{equation}
\label{eq:f_planeWaveExpansion}
f(x) = e^{N\ln[\cos(x)+i \alpha \,\sin(x)]}\approx e^{i N \alpha x} e^{N \alpha^2 x^2/2} .
\end{equation}
Therefore, the superoscillating Wigner distribution $W_{\Psi}$ contains interference
terms, equivalent to expression~(\ref{eq:interferenceterm}), proportional to $\cos
(\frac{2 p }{\hbar}\frac{N\Delta x}{2} \alpha) = \cos (p \frac{L}{\hbar} \alpha) $. Thus
for $W_{+}$, analogously to Zurek's scale~$a_Z$ in the compass state~Fig.~\ref{fig1}~{\bf(a)},
superoscillatory structures with $\alpha$-fold reduced length scales, Fig.~\ref{fig1}~{\bf(d)},
yielding areas on the scale of
\begin{equation}
\label{eq:sub-zurekscale}
a_{SO}\approx \frac{h/P}{\alpha} \times \frac{h/L}{\alpha} \approx \frac{a_{Z}}{\alpha^{2}}
\end{equation}
arise.

For these superoscillatory structures to show, the `overspill' from the two adjacent
squeezed states~$\Phi_{-1}$ and $\Phi_{+1}$ has to be so small that their Wigner
distributions obey
\begin{equation}
W_{\Phi_{-1}}(0,0) + W_{\Phi_{+1}}(0,0) \ll |W_{\Psi}(0,0)| .
\end{equation}

\section{Conclusion}

We remind the reader of the fact that a quantum wave function cannot be \emph{strictly}
``bandwidth-limited", simultaneously in position and momentum. Our Wigner distributions are
confined by a finite area $A$ in phase-space, yet they feature regular sub-Zurek scale
structures, in this sense they are superoscillating.

Zurek's compass states provide interferometric sensitivity at the Heisenberg limit. As
Fig.~\ref{fig1}~{\bf(c)} illustrates, our superoscillating states change, under tiny
displacements in $x$, $p$ and $t$, in essentially the same way as regu\-lar compass\-
states; they therefore do not per\-form better at the detection of small shifts or
rotations.

At this stage, the formation of structures below the Zurek scale using superoscillations
in phase-space is primarily a surprising curiosity.  Superoscillating regions are known to
be tiny in extent and amplitude~\cite{Berry_JPA06,Aharonov_JPA11,Kempf_JMP05}. This is why
our superoscillating states cannot show sensitivity below the Heisenberg limit.

We have shown, to paraphrase Zurek's statement cited above, that phase-space structures
are frequently as small as, but not generally smaller than $a_{Z}$. Yet, superoscillating
states can generate localized small patches with regular structures on very much smaller
scales.

An interesting open question raised by the existence of sub-Planck and sub-Zurek scale
phase-space structures concerns their potential effects on simulations. Possibly, grids
finer than commonly assumed for numerical calculations~\cite{Kosloff_JPC88} have to be
used.

%


\begin{thebibliography}{10}%
\makeatletter
\providecommand \@ifxundefined [1]{%
 \@ifx{#1\undefined}
}%
\providecommand \@ifnum [1]{%
 \ifnum #1\expandafter \@firstoftwo
 \else \expandafter \@secondoftwo
 \fi
}%
\providecommand \@ifx [1]{%
 \ifx #1\expandafter \@firstoftwo
 \else \expandafter \@secondoftwo
 \fi
}%
\providecommand \natexlab [1]{#1}%
\providecommand \enquote  [1]{``#1''}%
\providecommand \bibnamefont  [1]{#1}%
\providecommand \bibfnamefont [1]{#1}%
\providecommand \citenamefont [1]{#1}%
\providecommand \href@noop [0]{\@secondoftwo}%
\providecommand \href [0]{\begingroup \@sanitize@url \@href}%
\providecommand \@href[1]{\@@startlink{#1}\@@href}%
\providecommand \@@href[1]{\endgroup#1\@@endlink}%
\providecommand \@sanitize@url [0]{\catcode `\\12\catcode `\$12\catcode
  `\&12\catcode `\#12\catcode `\^12\catcode `\_12\catcode `\%12\relax}%
\providecommand \@@startlink[1]{}%
\providecommand \@@endlink[0]{}%
\providecommand \url  [0]{\begingroup\@sanitize@url \@url }%
\providecommand \@url [1]{\endgroup\@href {#1}{\urlprefix }}%
\providecommand \urlprefix  [0]{URL }%
\providecommand \Eprint [0]{\href }%
\providecommand \doibase [0]{http://dx.doi.org/}%
\providecommand \selectlanguage [0]{\@gobble}%
\providecommand \bibinfo  [0]{\@secondoftwo}%
\providecommand \bibfield  [0]{\@secondoftwo}%
\providecommand \translation [1]{[#1]}%
\providecommand \BibitemOpen [0]{}%
\providecommand \bibitemStop [0]{}%
\providecommand \bibitemNoStop [0]{.\EOS\space}%
\providecommand \EOS [0]{\spacefactor3000\relax}%
\providecommand \BibitemShut  [1]{\csname bibitem#1\endcsname}%
\let\auto@bib@innerbib\@empty
\bibitem [{\citenamefont {{Schleich}}(2001)}]{Schleich_01}%
  \BibitemOpen
  \bibfield  {author} {\bibinfo {author} {\bibfnamefont {W.~P.}\ \bibnamefont
  {{Schleich}}},\ }\href@noop {} {\emph {\bibinfo {title} {Quantum Optics in
  Phase Space}}}\ (\bibinfo  {publisher} {Wiley-VCH},\ \bibinfo {year}
  {2001})\BibitemShut {NoStop}%
\bibitem [{\citenamefont {Wigner}(1932)}]{Wigner_PR32}%
  \BibitemOpen
  \bibfield  {author} {\bibinfo {author} {\bibfnamefont {E.}~\bibnamefont
  {Wigner}},\ }\href {\doibase 10.1103/PhysRev.40.749} {\bibfield  {journal}
  {\bibinfo  {journal} {Phys. Rev.}\ }\textbf {\bibinfo {volume} {40}},\
  \bibinfo {pages} {749} (\bibinfo {year} {1932})}\BibitemShut {NoStop}%
\bibitem [{\citenamefont {Zurek}(2001)}]{Zurek_01}%
  \BibitemOpen
  \bibfield  {author} {\bibinfo {author} {\bibfnamefont {W.~H.}\ \bibnamefont
  {Zurek}},\ }\href {\doibase 10.1038/35089017} {\bibfield  {journal} {\bibinfo
   {journal} {Nature}\ }\textbf {\bibinfo {volume} {412}},\ \bibinfo {pages}
  {712} (\bibinfo {year} {2001})},\ \Eprint {http://arxiv.org/abs/0201118}
  {0201118} \BibitemShut {NoStop}%
\bibitem [{\citenamefont {Berry}(1994)}]{Berry_WS94}%
  \BibitemOpen
  \bibfield  {author} {\bibinfo {author} {\bibfnamefont {M.}~\bibnamefont
  {Berry}},\ }\href@noop {} {\emph {\bibinfo {title} {Quantum Coherence and
  Reality; in celebration of the 60th Birthday of Yakir Aharonov}}},\ edited
  by\ \bibinfo {editor} {\bibfnamefont {J.~S.}\ \bibnamefont {Anandan}}\ and\
  \bibinfo {editor} {\bibfnamefont {J.~L.}\ \bibnamefont {Safko}}\ (\bibinfo
  {publisher} {World Scientific},\ \bibinfo {address} {Singapore},\ \bibinfo
  {year} {1994})\ pp.\ \bibinfo {pages} {55--65}\BibitemShut {NoStop}%
\bibitem [{\citenamefont {Aharonov}\ \emph {et~al.}(1990)\citenamefont
  {Aharonov}, \citenamefont {Anandan}, \citenamefont {Popescu},\ and\
  \citenamefont {Vaidman}}]{Aharonov_PRL90}%
  \BibitemOpen
  \bibfield  {author} {\bibinfo {author} {\bibfnamefont {Y.}~\bibnamefont
  {Aharonov}}, \bibinfo {author} {\bibfnamefont {J.}~\bibnamefont {Anandan}},
  \bibinfo {author} {\bibfnamefont {S.}~\bibnamefont {Popescu}}, \ and\
  \bibinfo {author} {\bibfnamefont {L.}~\bibnamefont {Vaidman}},\ }\href
  {\doibase 10.1103/PhysRevLett.64.2965} {\bibfield  {journal} {\bibinfo
  {journal} {Phys. Rev. Lett.}\ }\textbf {\bibinfo {volume} {64}},\ \bibinfo
  {pages} {2965} (\bibinfo {year} {1990})}\BibitemShut {NoStop}%
\bibitem [{\citenamefont {Aharonov}\ \emph {et~al.}(2011)\citenamefont
  {Aharonov}, \citenamefont {Colombo}, \citenamefont {Sabadini}, \citenamefont
  {Struppa},\ and\ \citenamefont {Tollaksen}}]{Aharonov_JPA11}%
  \BibitemOpen
  \bibfield  {author} {\bibinfo {author} {\bibfnamefont {Y.}~\bibnamefont
  {Aharonov}}, \bibinfo {author} {\bibfnamefont {F.}~\bibnamefont {Colombo}},
  \bibinfo {author} {\bibfnamefont {I.}~\bibnamefont {Sabadini}}, \bibinfo
  {author} {\bibfnamefont {D.}~\bibnamefont {Struppa}}, \ and\ \bibinfo
  {author} {\bibfnamefont {J.}~\bibnamefont {Tollaksen}},\ }\href {\doibase
  10.1088/1751-8113/44/36/365304} {\bibfield  {journal} {\bibinfo  {journal}
  {J. Phys. A: Math. Theor.}\ }\textbf {\bibinfo {volume} {44}},\ \bibinfo
  {pages} {365304} (\bibinfo {year} {2011})}\BibitemShut {NoStop}%
\bibitem [{\citenamefont {Berry}\ and\ \citenamefont
  {Popescu}(2006)}]{Berry_JPA06}%
  \BibitemOpen
  \bibfield  {author} {\bibinfo {author} {\bibfnamefont {M.~V.}\ \bibnamefont
  {Berry}}\ and\ \bibinfo {author} {\bibfnamefont {S.}~\bibnamefont
  {Popescu}},\ }\href {\doibase 10.1088/0305-4470/39/22/011} {\bibfield
  {journal} {\bibinfo  {journal} {J. Phys. A: Math. Theor.}\ }\textbf {\bibinfo
  {volume} {39}},\ \bibinfo {pages} {6965} (\bibinfo {year}
  {2006})}\BibitemShut {NoStop}%
\bibitem [{\citenamefont {Calder}\ and\ \citenamefont
  {Kempf}(2005)}]{Kempf_JMP05}%
  \BibitemOpen
  \bibfield  {author} {\bibinfo {author} {\bibfnamefont {M.~S.}\ \bibnamefont
  {Calder}}\ and\ \bibinfo {author} {\bibfnamefont {A.}~\bibnamefont {Kempf}},\
  }\href {\doibase 10.1063/1.1825076} {\bibfield  {journal} {\bibinfo
  {journal} {J. Math. Phys.}\ }\textbf {\bibinfo {volume} {46}},\ \bibinfo
  {pages} {012101} (\bibinfo {year} {2005})}\BibitemShut {NoStop}%
\bibitem [{\citenamefont {Rogers}\ \emph {et~al.}(2012)\citenamefont {Rogers},
  \citenamefont {Lindberg}, \citenamefont {Roy}, \citenamefont {Savo},
  \citenamefont {Chad}, \citenamefont {Dennis},\ and\ \citenamefont
  {Zheludev}}]{Dennis_NATMat12}%
  \BibitemOpen
  \bibfield  {author} {\bibinfo {author} {\bibfnamefont {E.~T.}\ \bibnamefont
  {Rogers}}, \bibinfo {author} {\bibfnamefont {J.}~\bibnamefont {Lindberg}},
  \bibinfo {author} {\bibfnamefont {T.}~\bibnamefont {Roy}}, \bibinfo {author}
  {\bibfnamefont {S.}~\bibnamefont {Savo}}, \bibinfo {author} {\bibfnamefont
  {J.~E.}\ \bibnamefont {Chad}}, \bibinfo {author} {\bibfnamefont {M.~R.}\
  \bibnamefont {Dennis}}, \ and\ \bibinfo {author} {\bibfnamefont {N.~I.}\
  \bibnamefont {Zheludev}},\ }\href {\doibase 10.1038/nmat3280} {\bibfield
  {journal} {\bibinfo  {journal} {Nature Mat.}\ }\textbf {\bibinfo {volume}
  {11}},\ \bibinfo {pages} {432} (\bibinfo {year} {2012})}\BibitemShut
  {NoStop}%
\bibitem [{\citenamefont {Kosloff}(1988)}]{Kosloff_JPC88}%
  \BibitemOpen
  \bibfield  {author} {\bibinfo {author} {\bibfnamefont {R.}~\bibnamefont
  {Kosloff}},\ }\href {\doibase 10.1021/j100319a003} {\bibfield  {journal}
  {\bibinfo  {journal} {J. Chem. Phys.}\ }\textbf {\bibinfo {volume} {92}},\
  \bibinfo {pages} {2087} (\bibinfo {year} {1988})}\BibitemShut {NoStop}%
\end{thebibliography}


\end{document}